\documentclass[a4paper,11pt]{article}
\usepackage{authblk}
\usepackage{graphicx} 
\usepackage{cite} 
\usepackage{hyperref}
\usepackage{amsmath}
\usepackage{amssymb}
\usepackage{subfig}
\usepackage{multirow}
\usepackage{fancyhdr}
\usepackage{amsmath}

\usepackage{adjustbox}
\usepackage[headheight=14pt]{geometry}
\usepackage{epstopdf}

\title{A Study on the Triggering of Nucleonic Direct Urca Processes in Neutron Stars of Specific Masses and Their Hyperon Dependence}

\author[a,b]{Y. Xu}
\author[a]{Y. F. Shen}
\author[a]{Q. Yuan}
\author[a]{X. L. Huang}
\author[a] {Y. B. Wang}

\affil[a]{Changchun Observatory, National Astronomical Observatories, Chinese Academy of Sciences, Changchun 130117, China}
\affil[b]{School of Astronomy and Space Sciences, University of Chinese Academy of Sciences, Beijing 100049, China}

\date{09.04.2025}

\fancypagestyle{plain}{ 
    \fancyhf{}
    \fancyhead[C]{Submitted to Chinese Physics C}
}
\begin{document}

\maketitle
This work aims to analyze how hyperons affect neutrino radiation properties in nucleonic direct URCA processes, expecting to provide useful references for finding evidence of the existence of hyperons in astronomical observations. This analysis is carried out using the GM1 and NL3 parameter sets under the SU(6) and SU(3) flavor symmetries in the relativistic mean field theory framework. Combined with the inferred mass and radius values of PSRs J1231-1411, J0030+0451, and J0740+6620, our results show that nucleonic direct Urca processes are absent in PSR J1231-1411 due to momentum conservation violation. In hyperon-containing PSR J0030+0451 (NL3 parameter set), the nucleonic direct Urca processes involving $e^{-}$/ $\mu^{-}$ would occur. A large inferred mass span induces hyperon fraction variations, affecting neutrino emissivity. If the inferred mass of PSR J0030+0451 exceeds approximately 1.8 $M_{\odot}$, the neutrino luminosity of the nucleonic direct Urca processes under the SU(3) flavor symmetry remains nearly the same as that in npe$\mu$ matter, without depending on hyperons. However, it exhibits an obvious hyperon dependence under the SU(6) spin-flavor symmetry. For hyperon-containing J0740+6620, the nucleonic direct Urca processes under the SU(3) flavor symmetry in GM1 parameter set predicts faster neutrino luminosity decline with hyperonic fraction than npe$\mu$ matter, and under the SU(6) spin-flavor symmetry in NL3 parameter set it shows monotonic decreasing trend. The research shows that hyperonic fraction significantly affect the neutrino radiation properties of the nucleonic direct URCA processes in neutron stars. Different-mass pulsars (e.g., PSRs J1231-1411, J0030+0451, J0740+6620) exhibit the distinct nucleonic direct URCA processes behaviors, dependent on inferred masses/radii, parameter sets, and theoretical models.


\footnote{$^\ast$ Corresponding authors. E-mail: xuy@cho.ac.cn, huangxl@cho.ac.cn, wangyb@cho.ac.cn}
\footnote{This work was supported by the Development Project of
Science and Technology of Jilin Province (grant number: 20250102012JC) and the Special Project for the Theoretical Basic Research of Changchun Satellite Observatory, National Astronomical Observatories, Chinese Academy of Sciences (grant number: Y990000205).}

\section{Introduction}
Currently, global pulsar research has made significant progress: the Australia Telescope National Facility (ATNF) database has recorded over 3,700 certified pulsars \cite{Manchester2005}. Since China’s Five-hundred-meter Aperture Spherical Radio Telescope (FAST) came into operation \cite
{Nan2006}, it has discovered about one thousand new pulsars, surpassing the combined discoveries of all other telescopes during the same period \cite
{Han2021,Li2024,Su2024,Han2025}. However, in stark contrast to these observational breakthroughs, the core questions about the internal composition of neutron stars remain unresolved. Although observations clearly confirm the extreme compactness of neutron stars — with radii of $\sim$ 10 km and masses up to 2 $M_{\odot}$ — direct observational constraints on their core composition, such as whether nucleons undergo hyperonization, deconfinement phase transitions, or form meson condensates, are still lacking \cite{Sun2023,Huang2025}.
The presence of hyperons in neutron star cores stands as a pivotal scientific inquiry, whose resolution could revolutionize our understanding of dense matter physics and astrophysical phenomena. If confirmed, hyperons  would not only soften the equations of state of neutron stars — thereby reshaping theoretical models of stellar structure — but also affect particle composition distributions, thermal evolution dynamics, and observable signatures such as rotational behaviors and gravitational wave emissions. Against this backdrop, neutron star equations of state, as theoretical frameworks describing the pressure-density relationship, play an irreplaceable role in understanding their overall structure and macroscopic properties \cite{Wu2025}. By self-consistently linking microscopic nuclear forces to macroscopic astrophysical parameters (e.g., mass-radius relations), the equations of state provide a critical theoretical foundation for revealing neutron star interior compositions \cite{Miao2024,Parmar2025,Chen2025}. In recent years, mass and radius inferences for four pulsars—PSRs J0030+0451, J0740+6620, J0437-4715, and J1231-1411—from NASA’s Neutron Star Interior Composition Explorer (NICER) experiment have further constrainted neutron star equations of state\cite{Fonseca2021,Salmi2024,Choudhury2024,Vinciguerra2024,Hoogkamer2025,He2025}. 

The thermal evolution and cooling processes of neutron stars serve as crucial probes of their core physical properties, maintaining profound and multi-dimensional links to the equations of state \cite{Zhu2024,Tsang2024,Marino2024,Mendes2025,Scurto2025}. As a critical aspect in compact star evolution, the thermal evolution of neutron stars deeply reflects their internal matter states, microscopic interactions, and macroscopic energy transport mechanisms. Cooling processes not only record the initial high-temperature state after a supernova explosion but also continuously trace stellar structural evolution through mechanisms like neutrino radiation and photon emission. Notably, Potekhin et al. have conducted systematic studies on the cooling mechanisms of isolated and transient accreting neutron stars, suggesting the existence of neutron stars undergoing rapid cooling via baryonic direct Urca processes \cite{Potekhin2015,Potekhin2018,Potekhin2023}. 
 Marino et al. further proposed that the baryonic direct Urca processes may explain the anomalously low surface temperatures observed in young neutron stars such as PSR J0205+6449, PSR B2334+61, and CXOU J08524617 \cite{Marino2024}, which has drawn increasing attention to the role of nucleonic direct Urca processes during neutron star neutrino cooling stages. In this work, we primarily analyze the potential effects of hyperons on neutrino emission properties in nucleonic direct Urca processes within the relativistic mean field model framework by integrating astronomical observations. We attempt to provide potential reference methods for astronomical observations to search for evidence of hyperons in neutron stars.

The content of this work is structured as follows. In Section 2, we introduce the theoretical methods for deriving neutrino emissivity and luminosity in the nucleonic direct Urca processes within the framework of the relativistic mean field theory. In Section 3, we discuss the neutrino radiation properties of the nucleonic direct Urca processes, and investigate the potential impact of hyperons on three pulsars with known masses on the luminosity of the nucleonic direct Urca based on the selected model parameters. Section 4 presents the summary and conclusions.

\section{Theoretical Framework}
\subsection{The Relativistic Neutrino Emissivity}
The nucleonic direct Urca processes efficiently generate neutrinos via weak interactions among nucleons within the neutron star core, dominating the rapid cooling during the early neutrino-dominated era. The nucleonic direct Urca processes consist of beta decay and capture reactions \cite{Lattimer1991}. Neutron $\beta$ decay: a neutron decays into a proton, emitting an lepton ( $l=e^{-}$ and $\mu^{-}$ ) and an antilepton neutrino ( $\bar{\nu}_{l}$ ). Proton lepton capture: A proton captures an lepton, converting into a neutron and emitting a neutrino ( $\nu_{l}$ ). The specific forms are as follows:
\begin{gather}
A1: n\rightarrow p+l+\bar{\nu}_{l},\nonumber
A2: p + l\rightarrow n + \nu_{l}.\nonumber
\end{gather}

The two reactions persist in the neutron star core at densities exceeding the nuclear saturation density ( $n_{0}\approx2.7\times10^{14}$ g cm${^{-3}}$, corresponding to $n_{0}\approx 0.16 $ fm$^{-3}$ ), forming a "cooling channel" for neutrino radiation. Each reaction produces neutrinos carrying energies far exceeding the typical energy of thermal photons, which ensures that neutrino radiation dominates the early cooling stage \cite{Shapiro1986,Kaminker2001}. 
The total relativistic neutrino emissivity $Q$ per unit volume and time in both reactions (A1 and A2) can be expressed by the Fermi Golden Rule \cite{Leinson2002,Huang2015}, namely,
\begin{eqnarray}
&Q&=\frac{457\pi}{10080}G_{F}^{2}C^{2}T^{6}\Theta(p_{Fl}+p_{FN_{2}}-p_{FN_{1}})\nonumber
\\
&\times&\{C_{V}C_{A}((\varepsilon_{F_{1}}p_{Fl}^{2}+\varepsilon_{F_{2}}p_{Fl}^{2})-(\varepsilon_{F_{1}}-\varepsilon_{F_{2}})(p_{FN_{1}}^{2}-p_{FN_{2}}^{2}))\nonumber
\\
&+&2C_{A}^{2}m_{N_{1}}^{*}m_{N_{2}}^{*}\mu_{l}+(C_{V}^{2}+C_{A}^{2})(\mu_{l}(2\varepsilon_{F_{1}}\varepsilon_{F_{2}}-m_{N_{1}}^{*}m_{N_{2}}^{*})\nonumber
\\
&+&\varepsilon_{F_{1}}p_{Fl}^{2}-\frac{1}{2}(p_{FN_{1}}^{2}-p_{FN_{2}}^{2}+p_{Fl}^{2})(\varepsilon_{F_{1}}+\varepsilon_{F_{2}}))\}.
\end{eqnarray}
Here, the weak-coupling constant is $G_{F}=1.436\times10^{-49}$ erg cm$^{3}$, the Cabibbo factor $C=\cos\theta_{C}=0.973$, the threshold factor $\Theta(x)=1$ when $x\geq0$, the vector constant $C_{V}=1$, and the axial-vector constants $C_{A}\simeq1.26$. The nucleonic kinetic energy is given by $\varepsilon_{F}=\sqrt{m_{N}^{*2}+P_{FN}^{2}}$.

The neutrino luminosity of the nucleonic direct Urca processes, $L_{\nu}^{D}$, can be obtained by integrating the neutrino emissivity $Q$ over the volume element $dV= 4\pi r^{2}(1-\frac{2m}{r})^{-\frac{1}{2}} dr$ of a neutron star \cite{Gnedin1993}, namely
\begin{eqnarray}
L_{\nu}^{D}=\int_{0}^{R} Q(r) e^{2\Phi} dV.
\end{eqnarray}
Here, $e^{\Phi}=(1-\frac{2m}{r})^{\frac{1}{2}}$.

\subsection{Relativistic Mean Field Theory}
\label{sect:Obs}
Here we apply the relativistic mean field approximation that includes $\sigma$, $\omega$, $\rho$, $\sigma^{*}$ and $\phi$ mesons to describe the properties of the cores of neutron stars. The total Lagrangian density can be expressed as
\begin{equation}
\begin{aligned}
\mathcal{L} = {} & \sum_B \mathcal{L}_{B} + \sum_l \mathcal{L}_{l} + \mathcal{L}_{m} \\
= {} & \overline{\psi}_B [i\gamma_\mu \partial^\mu - (m_B - g_{\sigma B} \sigma - g_{\sigma^* B} \sigma^*) - g_{\rho B} \gamma_{\mu} \vec{\rho}^\mu \cdot \vec{I}_{B} - g_{\omega B} \gamma_\mu \omega^\mu \\
& - g_{\phi B} \gamma_\mu \phi^\mu] \psi_B + \overline{\psi}_l [i\gamma_\mu \partial^\mu - m_l] \psi_l - \frac{1}{4} W^{\mu v} W_{\mu v} - \frac{1}{4} \vec{R}^{\mu v} \cdot \vec{R}_{\mu v} \\
&- \frac{1}{4} P^{\mu v} P_{\mu v}+ \frac{1}{2} m_\omega^2 \omega_\mu \omega^\mu + \frac{1}{2} m_\rho^2 \vec{\rho}_\mu \cdot \vec{\rho}^\mu + \frac{1}{2} m^2_{\phi} \phi_\mu \phi^\mu - \frac{1}{3} a \sigma^{3} - \frac{1}{4} b \sigma^4 \\
&+\frac{1}{4} c_3 (\omega_\mu \omega^\mu)^2+ \frac{1}{2} (\partial_\mu \sigma \partial^\mu \sigma - m_\sigma^2 \sigma^2) + \frac{1}{2} (\partial_v \sigma^* \partial^v \sigma^* - m^2_{\sigma^*} \sigma^{*2}),
\end{aligned}
\end{equation}
with $\psi_{B(l)}$ the baryonic (leptonic) Dirac field, $\gamma_{u}$ the Dirac matrice, $\vec{I}_{B}$ the baryonic isospin matrix, $m_B(l)$ the baryonic (leptonic) mass. $g_{\omega B}$, $g_{\rho B}$ and $g_{\phi B}$ denote separately the $\omega-$, $\rho-$ and $\phi-B$ coupling constants. $W_{\mu v}=\partial_\mu\omega_v-\partial_v\omega_\mu$, $R_{\mu v}=\partial_\mu{\mathbf{\rho}}_v-\partial_v{\mathbf{\rho}}_\mu$ and $P_{\mu v}=\partial_\mu\phi_v-\partial_v\phi_\mu$ correspond individually to the field strength tensors for the $\omega$, $\rho$ and $\phi$ mesons. The sum B runs over N (neutron and proton), $\Lambda$, $\Xi^{-}$ and $\Xi^{0}$. We have not considered $\Sigma^{+}$, $\Sigma^{0}$, $\Sigma^{-}$ hyperons due to the potential uncertainty of $\Sigma$ hyperons at saturation density in nuclear matter \cite{Batty1994,Batty2005,Yu2011}.

In the relativistic mean field approximation, the five meson fields are replaced by their expectation values ( $\sigma^{0}, \sigma^{*0}, \omega^{0}, \rho^{0}_{3}$ and $\phi^0$ ) which yields the following linear Dirac equation for the baryonic field,
\begin{eqnarray}
(i\gamma_{\mu}\partial^{\mu}-m_B^{*}-g_{\omega B}\gamma_{0}\omega^{0}-g_{\rho B}\gamma_{0}I_{3B}\rho^{0}_{3}-g_{\phi B}\gamma_{0}\phi^0)\psi_{B}=0,
\end{eqnarray}
where $I_{3B}$ is the third component of the baryonic isospin. $m_B^{*}$ denotes the baryonic effective mass and can be written as
$m_B^{*}=m_B-g_{\sigma B}\sigma-g_{\sigma^*B}\sigma^*$.
$g_{\sigma B}$ and $g_{ \sigma^{*} B}$ are the $\sigma-$ and $\sigma^{*}-B$ coupling constants, respectively. 

Thus, the five average meson fields are solutions of the following equations of motion 
\begin{eqnarray}
\label{eq: baryon and meson fields}
\sum_B \frac{g_{\sigma B}}{\pi^{2}}\int_0^{p_{FB}}\frac{m_B^{*}p_{B}^{2}dp_{B}}{(m_B^{*2}+p_{B}^{2})^{1/2}}=m_\sigma^2\sigma^0+a(\sigma^{0})^2+b(\sigma^0)^3,\\
\sum_B \frac{g_{\sigma^* B}}{\pi^{2}}\int_0^{p_{FB}}\frac{m_B^{*}p_{B}^{2}dp_{B}}{(m_B^{*2}+p_{B}^{2})^{1/2}}=m_{\sigma^{*0}}^2\sigma^{*0},\\
\sum_B g_{\omega B}\frac{p_{FB}^{3}}{3\pi^2}=m_\omega^2\omega^0+c_{3}(\omega^{0})^{3},\\
\sum_B g_{\rho B}\frac{p_{FB}^{3}}{3\pi^2}I_{3B}=m_{\rho}^2\rho^{0}_{3},\\
\sum_B g_{\phi B}\frac{p_{FB}^{3}}{3\pi^2}=m_\phi^2\phi^0.
\end{eqnarray}
$p_{FB (l)}$ denotes the baryonic (leptonic) Fermi momentum. 

The baryonic and leptonic chemical potentials are given by
\begin{eqnarray}
\mu_{B}=\sqrt{p_{FB}^2+{m _{B}^*}^2}+g_{\omega B}+g_{\phi B}\phi^0
\omega_{0}+g_{\rho B}\rho_{0} I_{3B},
\\
\mu_{l}=\sqrt{p_{Fl}^2+m_{l}^2}.
\end{eqnarray}
We can derive the equations of state under the conditions of electric neutrality and beta equilibrium in neutron stars. We can thereby obtain the required nucleonic effective mass, chemical potential, and Fermi momentum, as well as the leptonic chemical potential and Fermi momentum, for calculating the neutrino emissivity in the nucleonic direct Urca processes (as shown in Eq. 1). By solving the Tolman-Oppenheimer-Volkoff equation \cite{Oppenheimer1939,Tolman1939}, the mass-radius relationships of neutron stars can be further obtained. Moreover, the Keplerian frequency can be obtained from the Relativistic Roche model \cite{Shapiro1989,Haensel2009}, and its relationship with the mass and radius of a neutron star is expressed as
$f_{\mathrm{max}}^{\mathrm{Roche}} \approx 1.0~\mathrm{kHz} (\frac{M}{M_\odot})^{\frac{1}{2}}(\frac{R}{10~\mathrm{km}})^{-\frac{3}{2}}$.    

\begin{table*}
    \caption{Threshold density and mass values for the emergence of $\Lambda$, $\Xi^{-}$ and $\Xi^{0}$ hyperons for three cases SU3: npe$\mu$+hyperons, SU6: npe$\mu$+hyperons and SU6-no $\sigma^{*}$, $\phi$: npe$\mu$+hyperons.} 
    \label{tab:our data}
    \centering
    \small 
    \setlength{\tabcolsep}{1pt} 
    \begin{tabular}{l|l|cc|cc|cc}\hline
 Parameter Set & Hyperon & \multicolumn{2}{c}{case(ii)} & \multicolumn{2}{c}{case(iii)} & \multicolumn{2}{c}{case(iv)} \\
  \cline{3-8}
  && $n_B$ (fm$^{-3}$) & $M_c(M_\odot)$ & $n_B$ (fm$^{-3}$) & $M_c(M_\odot)$ & $n_B$ (fm$^{-3}$) & $M_c(M_\odot)$ \\
	\hline
GM1 &$\Lambda$ &0.386 &1.671&0.349&1.502&0.349&1.502  \\
    & $\Xi^{-}$&0.459 &1.888&0.398&1.653&0.396&1.644  \\
    & $\Xi^{0}$&0.791 &2.136&0.704&1.856&0.730&1.819  \\
    \hline
NL3	&$\Lambda$ &0.303 &1.702&0.285&1.538&0.285&1.538   \\
    &$\Xi^{-}$ &0.360 &2.023&0.334&1.767&0.334&1.759   \\
    &$\Xi^{0}$ &0.575 &2.403&0.567&2.042&0.590&1.989   \\
\hline
    \end{tabular}
\end{table*}

\begin{figure*}
    \centering
    \adjustbox{scale=0.89}{ 
        \begin{tabular}{cc}
            \subfloat[][\label{f2:a}]{\includegraphics[width=0.45\linewidth]{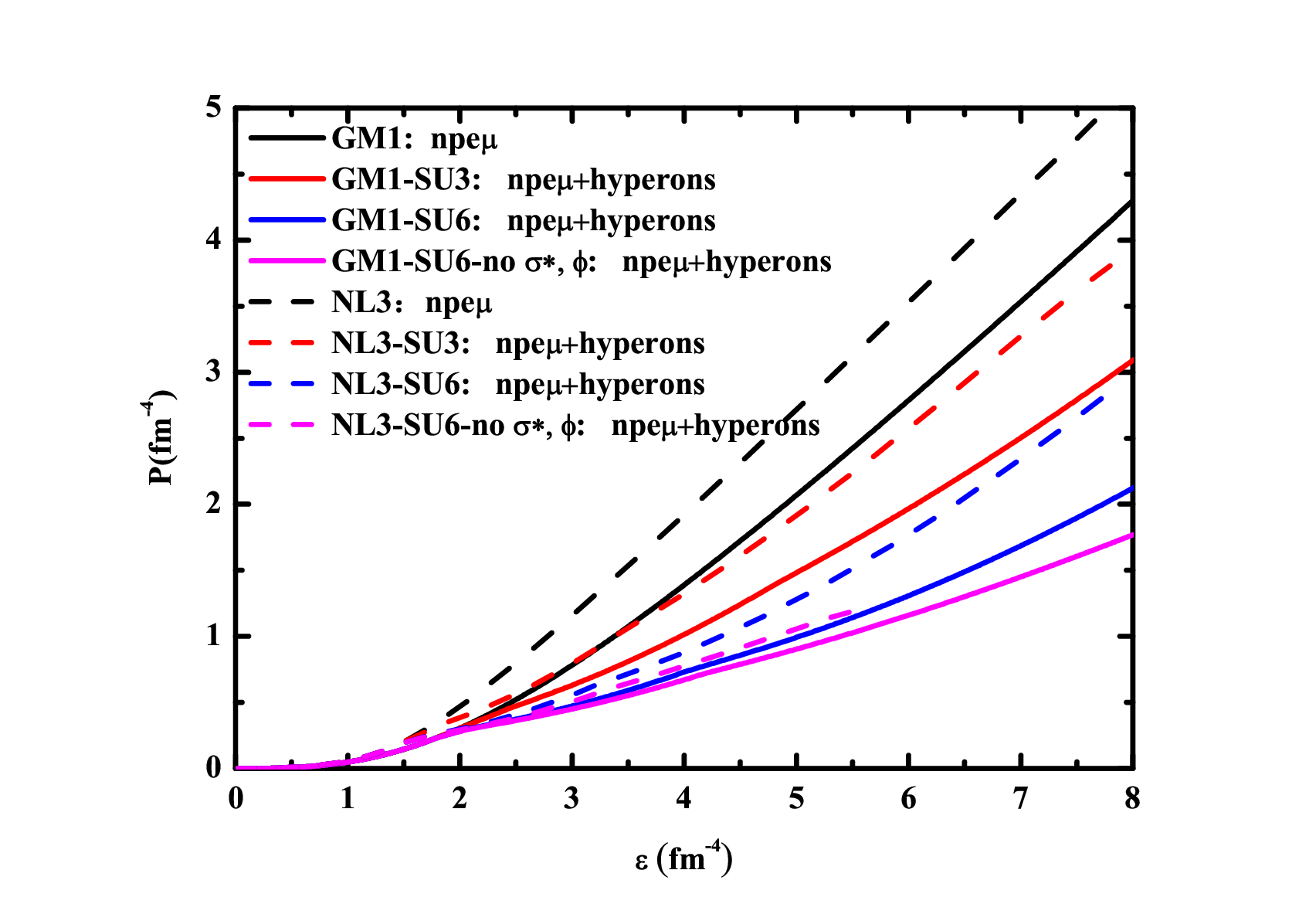}} &
            \subfloat[][\label{f2:b}]{\includegraphics[width=0.45\linewidth]{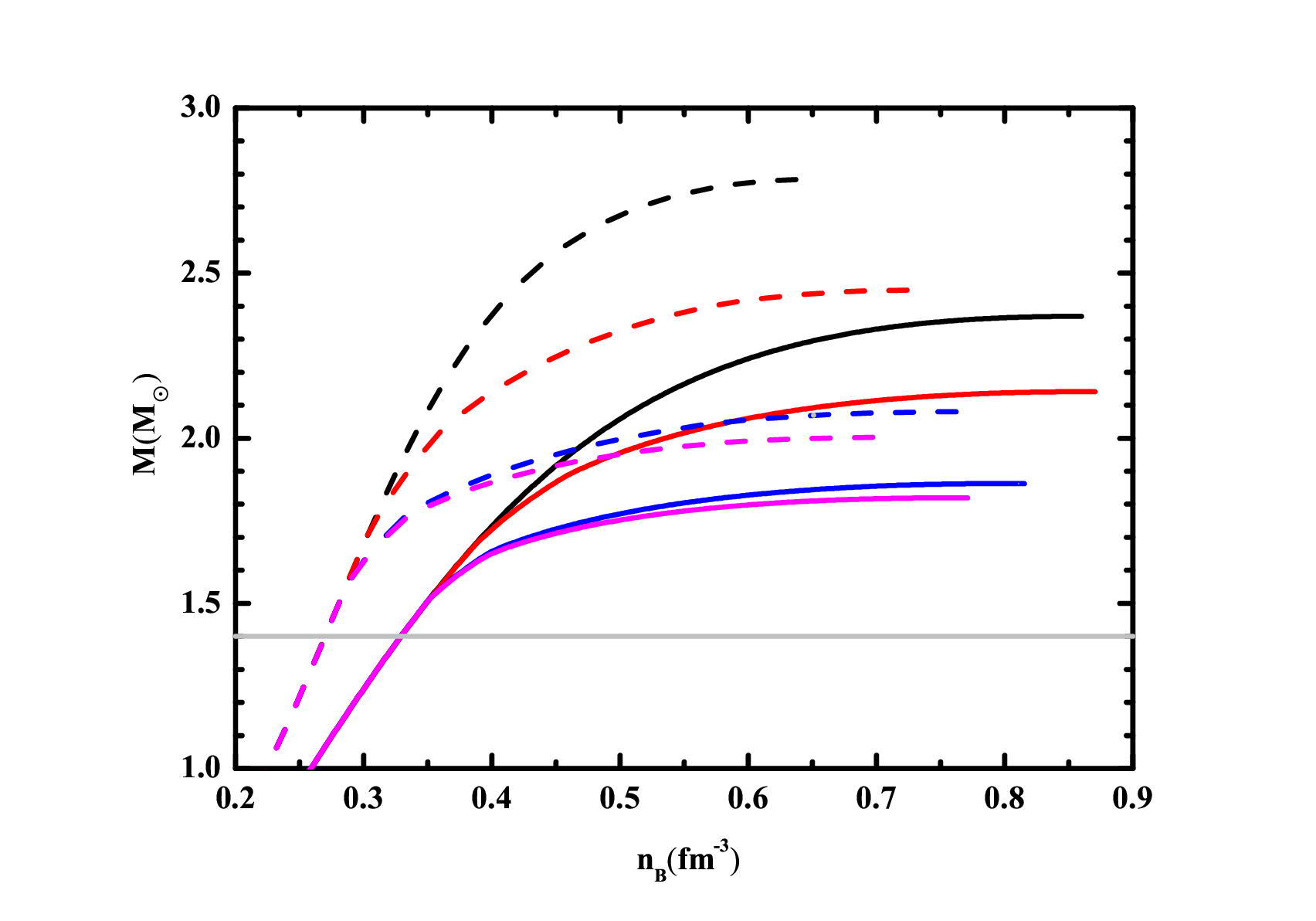}} \\
            \subfloat[][\label{f2:c}]{\includegraphics[width=0.45\linewidth]{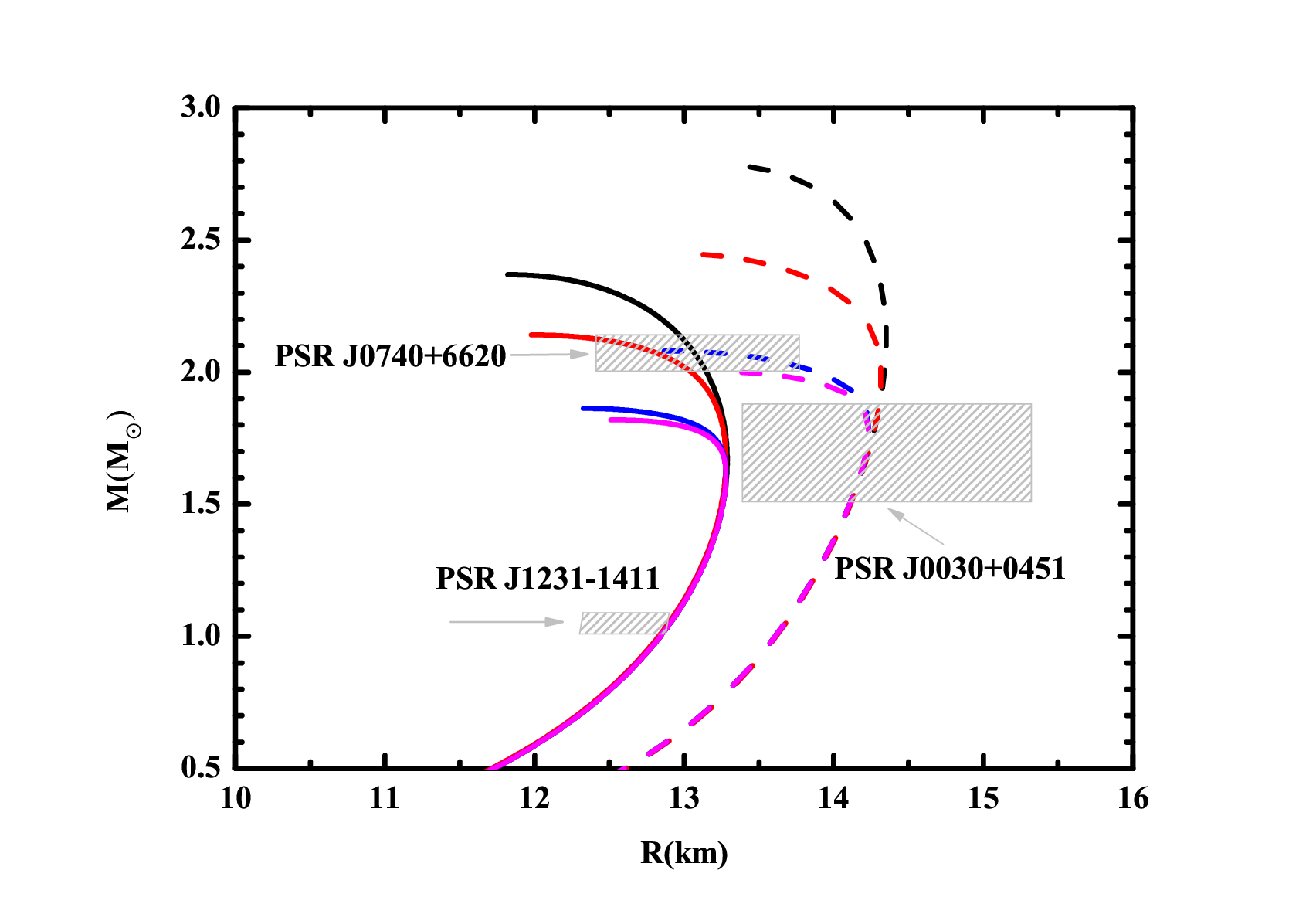}} &
            \subfloat[][\label{f2:d}]{\includegraphics[width=0.45\linewidth]{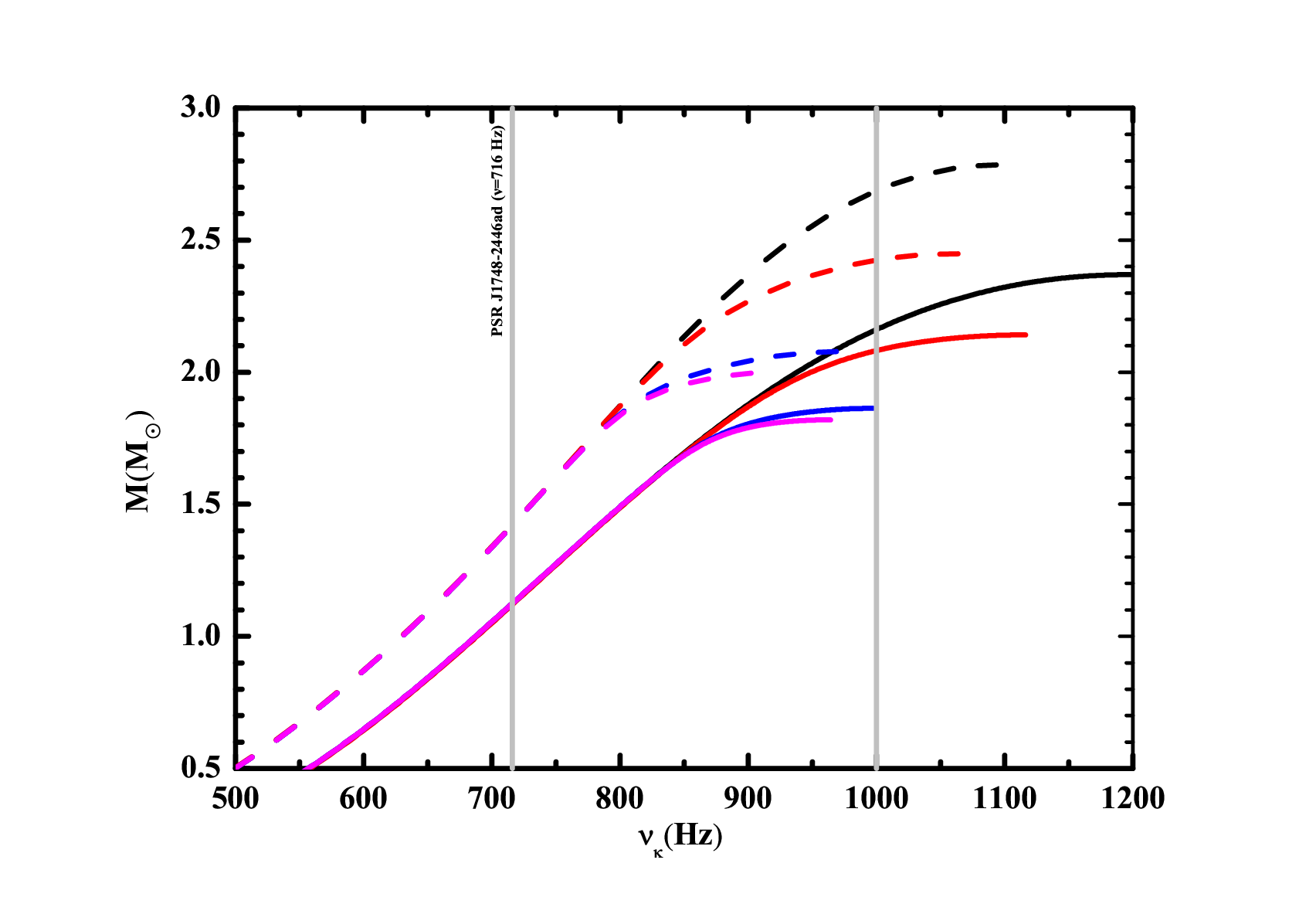}}
        \end{tabular}
    }
    \caption{The equations of state, mass-density, mass-radius, and mass-frequency relations of neutron stars in four cases of GM1 and NL3 parameter sets.}
    \label{fig:eos mass-radius mass-frequency relations}
\end{figure*}

\section{Analysis and Discussion}
In the work, we analyzed the influence of hyperons on the mass-radius, mass-density, mass-Kepler frequency, neutrino emissivity-radius, and neutrino luminosity-mass relationships of neutron stars which are obtained by the relativistic mean field theory model with the parameter sets GM1 and NL3. To this end, we considered four different equations of state:
(i) A neutron star consists of only neutrons, protons, electrons and muons, denoted as npe$\mu$. (ii) A neutron star contains neutrons, protons, electrons, muons and hyperons ($\Lambda$, $\Xi^{-}$ and $\Xi^{0}$). Regarding the baryon-meson coupling constants, five mesons, namely $\sigma$, $\omega$, $\rho$, $\sigma^{*}$ and $\phi$, are included in the general SU(3) flavor symmetry. This case is denoted as SU3: npe$\mu$+hyperons. (iii) It is the same as case (ii), except that the hyperon-meson coupling constats in this case follow the SU(6) spin-ﬂavor symmetry. Case (iii) is named SU6: npe$\mu$+hyperons. (iv) It is identical to case (iii), but without the $\sigma^{*}$ and $\phi$ mesons, and is denoted as SU6-no $\sigma^{*}$, $\phi$: npe$\mu$+hyperons.
 
The possible equations of state, as well as the mass - radius, mass - density, and mass - Kepler frequency relationships for the above four cases under the GM1 and NL3 parameter sets, are shown in Figure \ref{fig:eos mass-radius mass-frequency relations}. Additionally, table 1 lists the threshold densities and mass values for the emergence of hyperons in cases (ii - iv) with the GM1 and NL3 parameter sets. Figure \ref{fig:eos mass-radius mass-frequency relations} shows that the equations of state of neutron star matter soften significantly when hyperons are present.
Notably, in case (iv), under the SU(6) spin - flavor symmetry, it results in the softest equations of state and thus it corresponds to the smallest neutron star mass among the four cases. In Figure \ref{f2:c}, the shaded regions represent the inferred mass and radius values of PSRs J1231-1411 ($1.04^{+0.05}_{-0.03}$ $M_{\odot}$ with radius $12.6\pm0.3$ km), J0030+0451 ($1.70^{+0.18}_{-0.19}$ $M_{\odot}$ with radius $14.44^{+0.88}_{-1.05}$ km) and J0740+6620 ($2.073^{+0.069}_{-0.069}$ $M_{\odot}$ with radius $12.49^{+1.28}_{-0.08}$ km), respectively. 
Based on Figure \ref{f2:c} and Table 1, the inferred mass and radius values of PSR J1231-1411 align with the mass - radius relations of the GM1 parameter set, indicating that it is only likely to be a low-mass conventional neutron star within this theoretical framework. If the factor of hyperons is not taken into account, PSRs J0030+0451 and J0740+6620 are very likely to belong to the intermediate-mass and the massive traditional neutron stars, respectively. When hyperons are present in the interior of a neutron star, the inferred mass and radius values of PSR J0031+0451 exclusively match the mass - radius relationships of the NL3 parameter set. This suggests that it may be classified as an intermediate - mass hyperon star. Its interior will only have $\Lambda$ hyperons under the SU3 flavor symmetry, while  it will have either $\Lambda$ hyperons or both $\Lambda$ and $\Xi^{-}$ hyperons under the SU6 spin-flavor symmetry. As for PSR J0740+6620, its inferred mass and radius values are consistent with the mass - radius relations of case (ii) under the GM1 parameter set and case (iii) under the NL3 parameter set. This strongly indicates that this pulsar is likely a massive hyperon star, with its interior containing $\Lambda$, $\Xi^{-}$, or $\Lambda$, $\Xi^{-}$ and $\Xi^{0}$ hyperons. Figure \ref{f2:d} presents the mass - frequency relationships for the GM1 and NL3 parameter sets under both symmetry assumptions, incorporating the observational constraints from the currently known fastest - spinning pulsar: PSR J1748 - 2446ad with a spin frequency of $\nu$ = 716 Hz \cite{Hessels2006}. Under our model and parameter settings, PSR J1748 - 2446ad is likely a conventional neutron star with a relatively small mass. As shown in Figure \ref{f2:d}, when the Keplerian frequency exceeds 1000 Hz, the possible mass range of neutron stars varies significantly. According to the calculation results of the GM1 and NL3 parameter sets, the mass range of sub-millisecond pulsars with $f=$ 1000 Hz is approximately between 1.86 $M_{\odot}$ and 2.70 $M_{\odot}$, falling within the category of medium to high mass.
\begin{figure*}[htbp]
    \centering
    \adjustbox{scale=0.89}{ 
        \begin{tabular}{cc}
            \subfloat[][\label{f3:a}]{\includegraphics[width=0.45\linewidth]{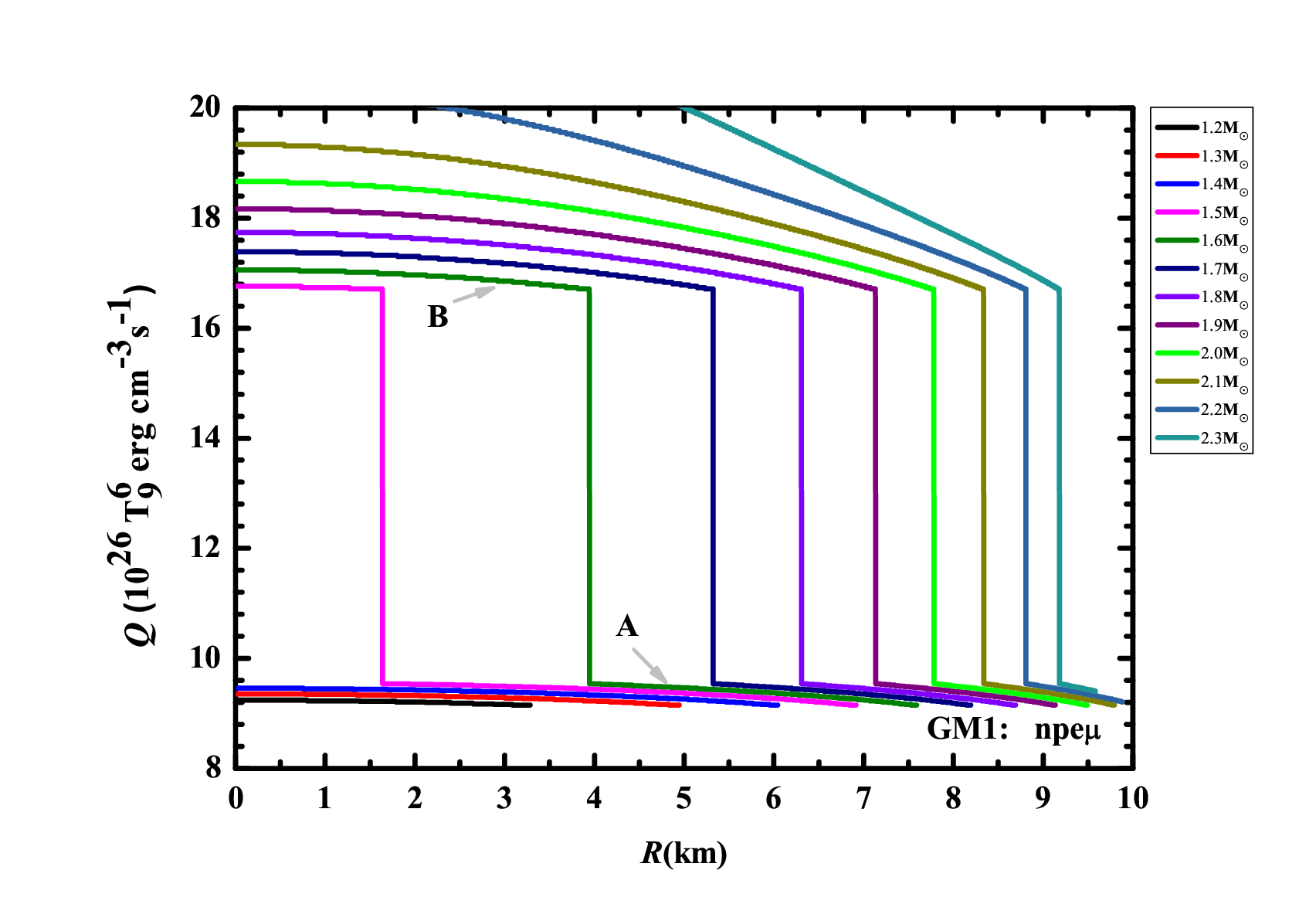}} &
            \subfloat[][\label{f3:b}]{\includegraphics[width=0.45\linewidth]{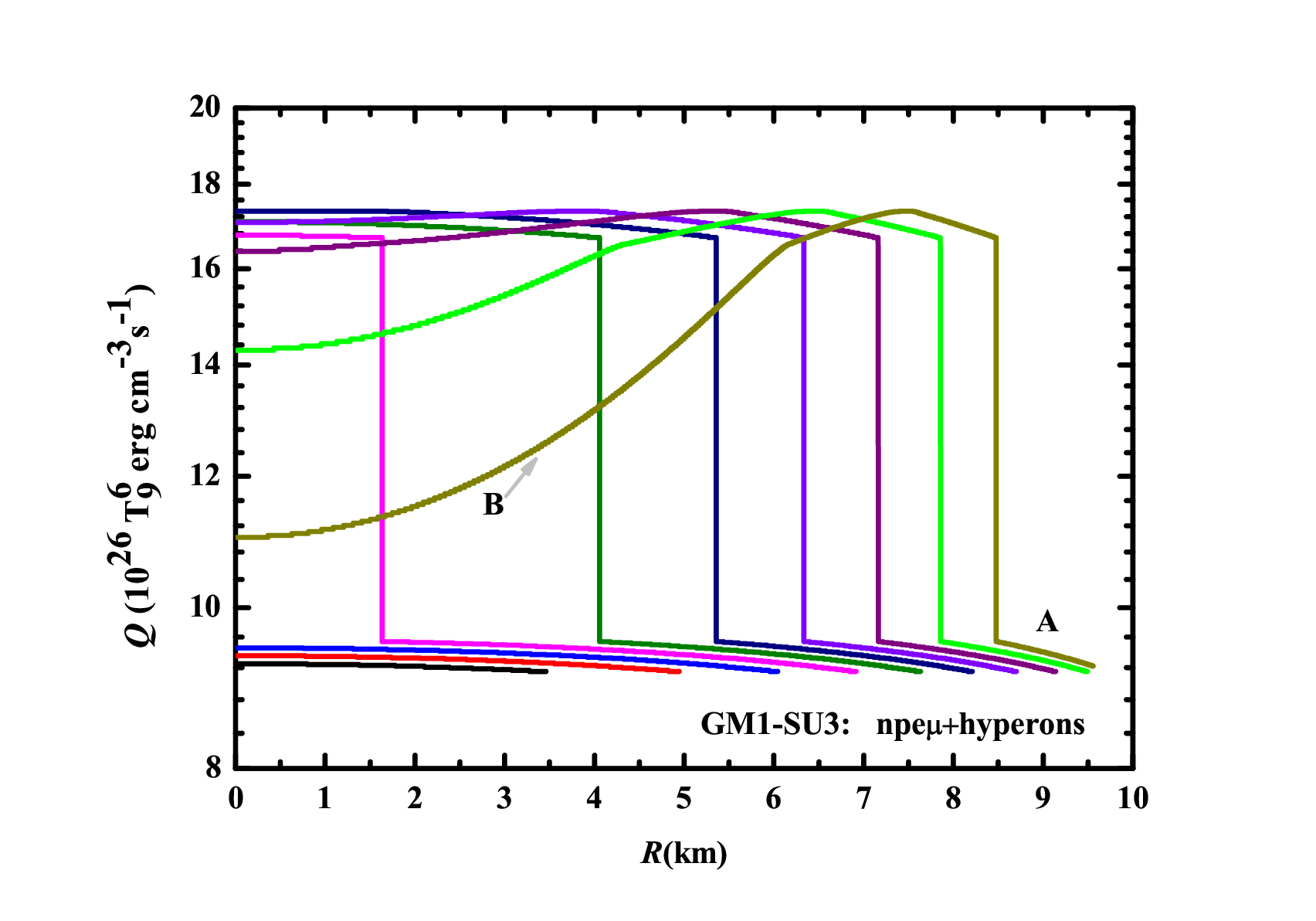}} \\
            \subfloat[][\label{f3:c}]{\includegraphics[width=0.45\linewidth]{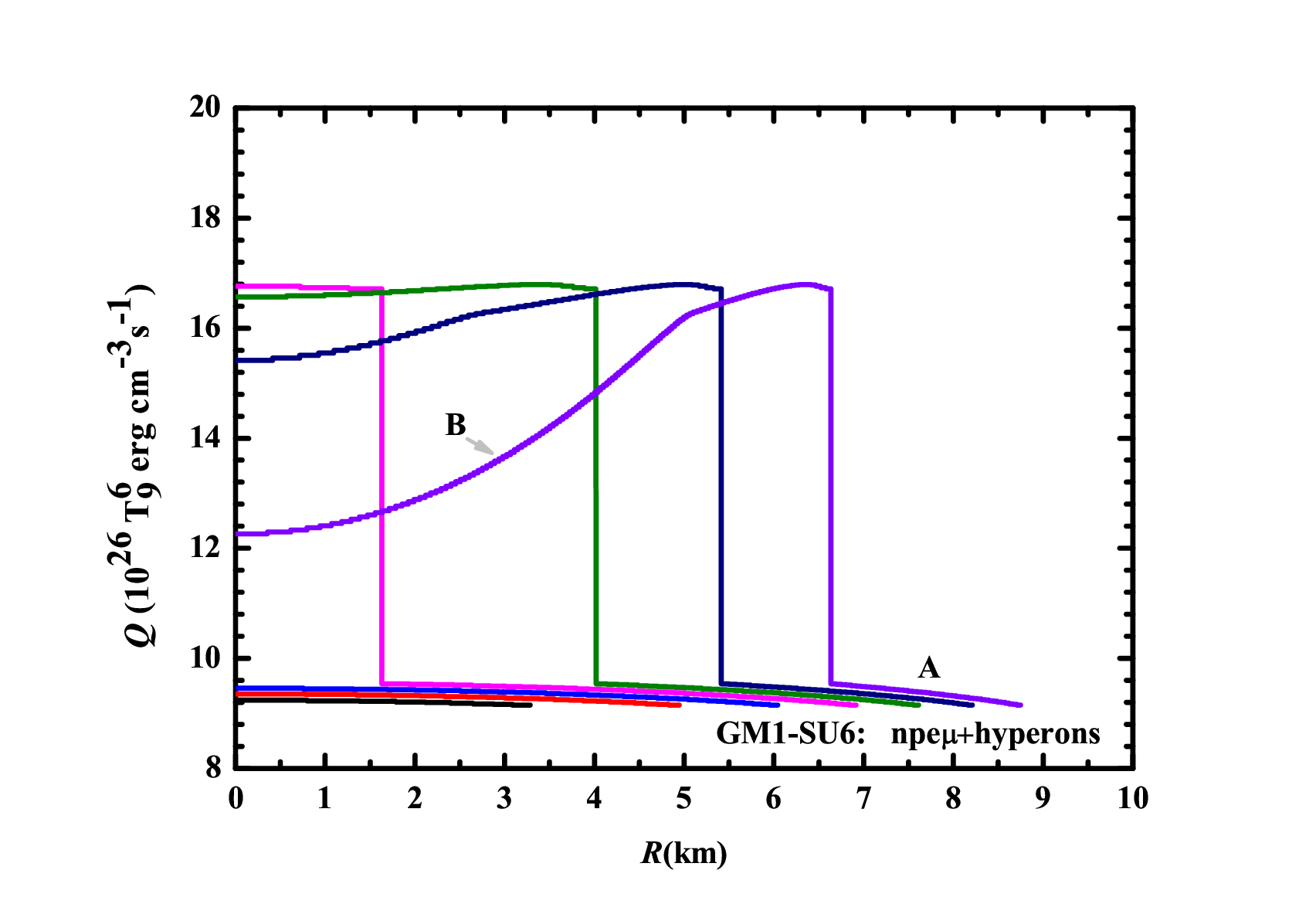}} &
            \subfloat[][\label{f3:d}]{\includegraphics[width=0.45\linewidth]{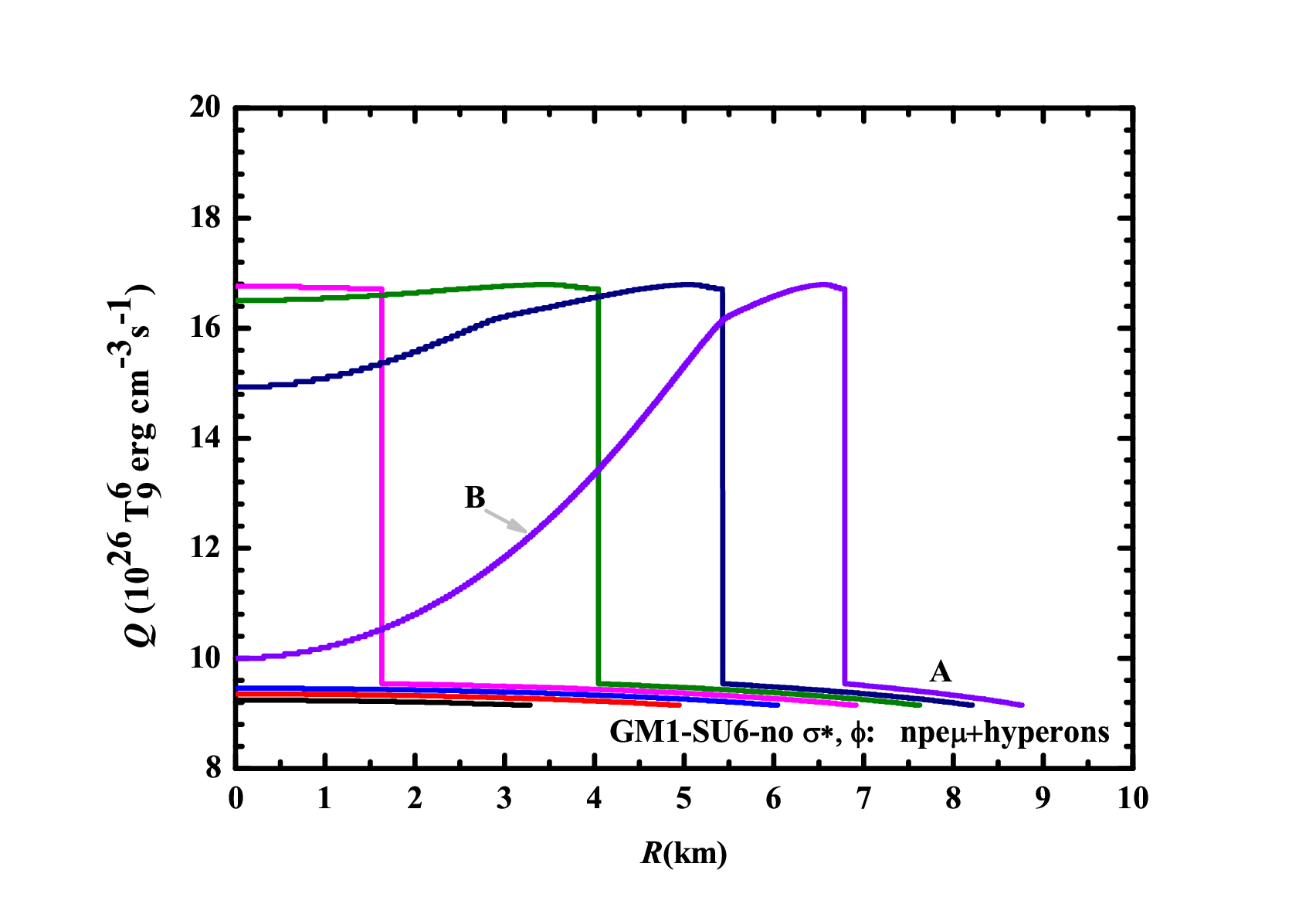}}
        \end{tabular}
    }
    \caption{Radial variation diagrams of the neutrino emissivity in the nucleonic direct Urca processes within the core of neutron stars for four cases of the GM1 parameter set.}
    \label{fig:neutrino emissivity GM1}
\end{figure*}

\begin{figure*}[htbp]
    \centering
    \adjustbox{scale=0.89}{
        \begin{tabular}{cc}
            \subfloat[][\label{f4:a}]{\includegraphics[width=0.45\linewidth]{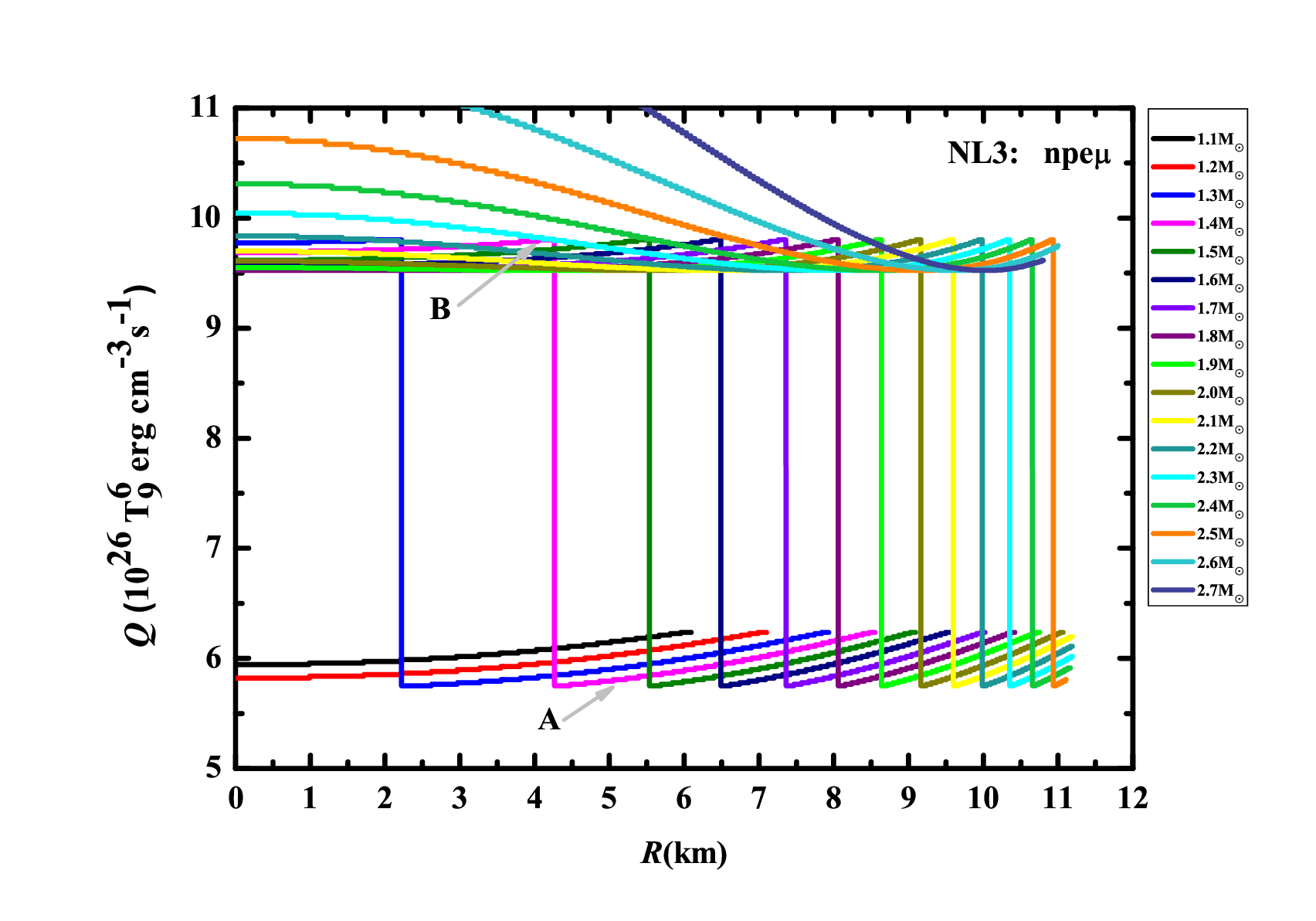}} &
            \subfloat[][\label{f4:b}]{\includegraphics[width=0.45\linewidth]{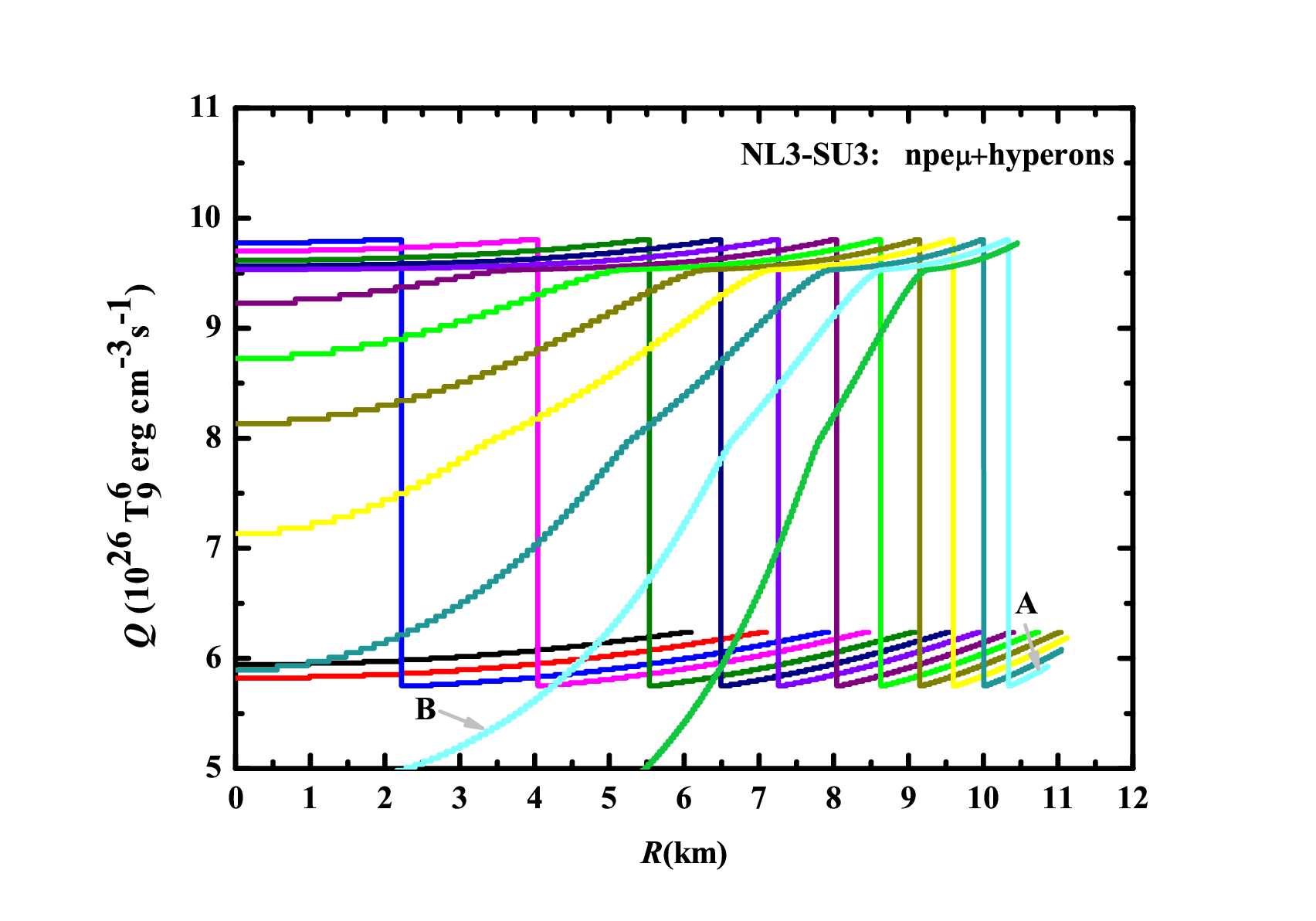}} \\
            \subfloat[][\label{f4:c}]{\includegraphics[width=0.45\linewidth]{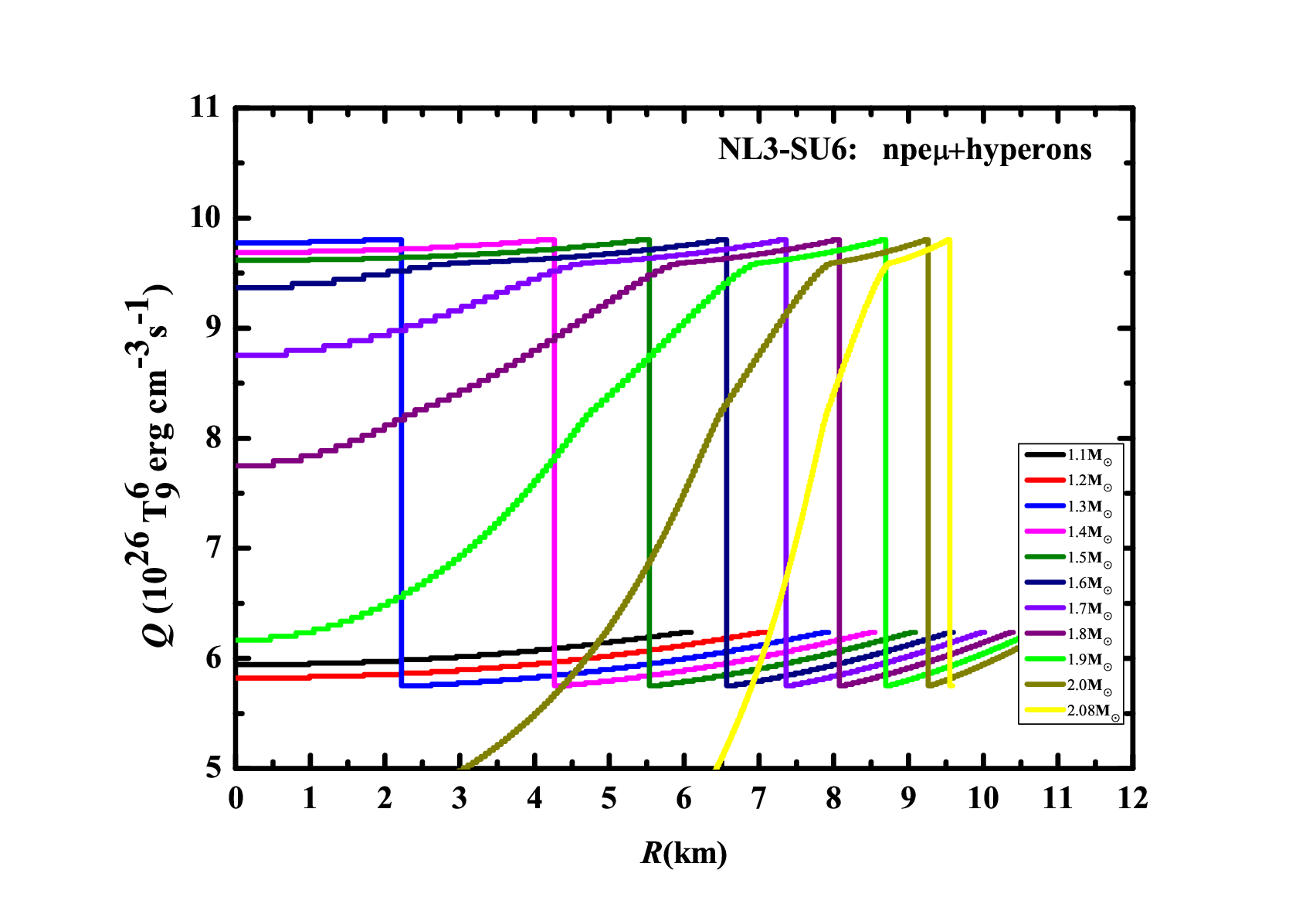}} &
            \subfloat[][\label{f4:d}]{\includegraphics[width=0.45\linewidth]{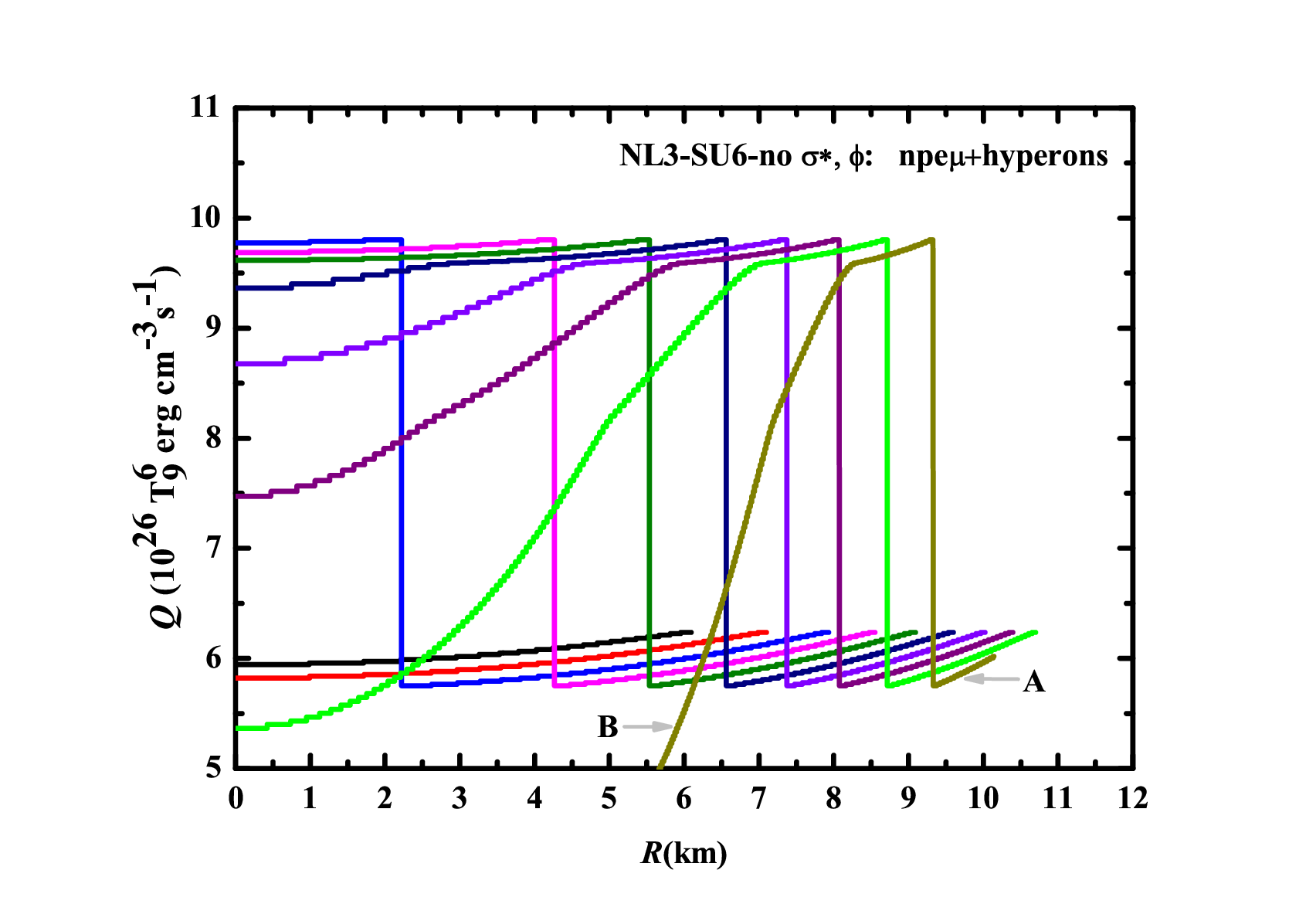}}
        \end{tabular}
    }
    \caption{Radial variation diagrams of the neutrino emissivity in the nucleonic direct Urca processes within the core of neutron stars for four cases of the NL3 parameter set.}
    \label{fig:neutrino emissivity NL3}
\end{figure*}

The neutrino emissivity of the nucleonic direct Urca processes as the function of stellar radius for the four cases under the GM1 and TM1 parameter sets is presented in Figures \ref{fig:neutrino emissivity GM1} and \ref{fig:neutrino emissivity NL3}, respectively. It should be noted that each curve comprises two distinct components (labeled A and B). Specifically, $e^{-}$ is present in the entire interior of a neutron star, whereas $\mu^{-}$ requires the higher threshold densities 0.129 fm$^{-3}$ (GM1) and 0.111 fm$^{-3}$ (NL3), respectively. Consequently, component A stands for the sole contribution of the nucleonic direct Urca processes with $e^{-}$ to neutrino emissivity, while component B incorporates the combined contributions from the nucleonic direct Urca processes with both $e^{-}$ and $\mu^{-}$ to neutrino emissivity. It is shown that only the nucleonic direct Urca processes with $e^{-}$ occur in the neutron star with masses spanning 1.2 $M_{\odot}$–1.4 $M_{\odot}$ within the GM1 equations of state framework( 1.1 $M_{\odot}$–1.2 $M_{\odot}$ under NL3 parameters ), maintaining a relatively stable neutrino emissivity. However, when the center density of the neutron star approaches the threshold density of $\mu^{-}$, a significant enhancement in neutrino emissivity arises from the nucleonic direct Urca processes involving $\mu^{-}$, triggering an abrupt rise in the total neutrino emissivity. Furthermore, it might illustrate that there is an existence of an inverse correlation between the hyperonic fraction and the neutrino emissivity of B section by the comprehensive analysis of the hyperon threshold densities and critical masses across four cases in Table 1 and Figures \ref{fig:neutrino emissivity GM1} and \ref{fig:neutrino emissivity NL3}, where the greater hyperonic fraction induces the stronger suppression of the neutrino emissivity for the nucleonic direct Urca processes. Because the threshold mass required for the appearance of hyperons under the SU(3) flavor symmetry is higher than that under the SU(6) spin-flavor symmetry, we will take the change in the neutrino emissivity within a neutron star with a mass of 1.8 $M_{\odot}$ as an example to illustrate it in detail. In case (ii), the hyperonic fraction inside this hyperon star with a mass of 1.8 $M_{\odot}$ is relatively low. As a result, there is almost no difference in the intensity of neutrino energy loss during the nucleonic direct Urca processes in their cores for a traditional neutron star and a hyperon star both with a mass of 1.8 solar masses. On the contrary, in cases (iii) and (iv), the threshold mass required for the appearance of hyperons is reduced, which significantly increases the hyperonic fraction inside the hyperon star with a mass of 1.8 $M_{\odot}$. Therefore, there is a significant difference in the intensity of neutrino energy loss during the nucleonic direct Urca processes in their cores for a traditional neutron star and a hyperon star both with a mass of 1.8 $M_{\odot}$, and hyperons will obviously suppress the neutrino emission intensity of the hyperon star with a mass of 1.8 $M_{\odot}$.
In addition, since the nucleonic and leptonic Fermi momenta within PSR J1231-1411 do not satisfy the momentum conservation condition in Eq.1, the nucleonic direct Urca processes are prevented from occurring within this pulsar. According to the numerical results of the GM1 or NL3 parameter set, a canonical neutron star with a mass of 1.4 
$M_{\odot}$ is categorized as a traditional neutron star, and its neutrino emissivity for the nucleonic direct Urca processes is found to between (9.154 - 9.457) $\times$ $10^{26}$ T$^{6}_{9}$ erg cm$^{-3}$s$^{-1}$ or (5.752 - 9.803) 
$\times$ $10^{26}$ T$^{6}_{9}$ erg cm$^{-3}$s$^{-1}$.
Because the inferred mass range of PSR J0030+0451 ( 1.51 $M_{\odot}$ - 1.88 $M_{\odot}$ ) has a large span, it gives rise to obvious fluctuations in the hyperonic fraction. As depicted in Figure \ref{fig:neutrino emissivity NL3}, this leads to significant changes in the intensity of the neutrino emissivity of the nucleonic direct Urca processes, with the maximum change occurring in case (iv). Similarly, the inferred mass range of PSR J0740+6620 ( 2.004 $M_{\odot}$ - 2.142 $M_{\odot}$ ) has a smaller span compared to that of PSR J0030+0451. Therefore, the fluctuations of the hyperonic fraction are also smaller. Figure \ref{fig:neutrino emissivity GM1} and Figure \ref{fig:neutrino emissivity NL3} indicate that the change in the intensity of the neutrino emissivity of the nucleonic direct Urca processes is also relatively small.

\begin{table*}
   \caption{Neutrino luminosity values and ranges for the  nucleonic direct Urca processes in PSRs J0030+0451 and J0740+6620 under constraints of the inferred mass $M_{mv}$, the inferred central mass $M_{mv}^{c}$, the inferred compactness $(M/R)_{mv}$, and the inferred central compactness $(M/R)_{mv}^{c}$ for the GM1 and NL3 parameter sets.}
    \label{tab:neutrino_luminosity_values}
    \centering 
     \small 
    \setlength{\tabcolsep}{6pt} 
    \begin{tabular}{c|c|cc} 
	\hline\hline 
\multirow{2}{*}{Parameter Set} & \multirow{2}{*}{Physical quantities} & \multicolumn{2}{c}{$L_{\nu}^{D}$ ($10^{45}$ T$_{9}^{6}$ erg s$^{-1}$)} \\
	\cline{3-4} 
	& & PSR J0030+0451 & PSR J0740+6620 \\
	\hline 
	\multirow{4}{*}{GM1} 
	    & $M_{mv}$        & -- & 3.376--5.155 \\
	    & $(M/R)_{mv}$    & -- & 3.879--5.307 \\
	    & $M_{mv}^{c}$     & -- & 4.449--4.795 \\
	    & $(M/R)_{mv}^{c}$ & -- & 4.359--5.193 \\
	\hline
	\multirow{4}{*}{NL3} 
	    & $M_{mv}$        & 2.068--3.573 & 1.962--3.265 \\
    & $(M/R)_{mv}$    & 1.550--4.027 & 1.962--3.178 \\
	    & $M_{mv}^{c}$     & 2.869        & 2.407 \\
	    & $(M/R)_{mv}^{c}$ & 2.760        & -- \\
	\hline\hline 
    \end{tabular}
    \vspace{0.3cm}
    \footnotesize{}
\end{table*}

\begin{figure*}[htbp]
    \centering
    \adjustbox{scale=0.89}{
        \begin{tabular}{cc}
            \subfloat[][\label{f5:a}]{\includegraphics[width=0.45\linewidth]{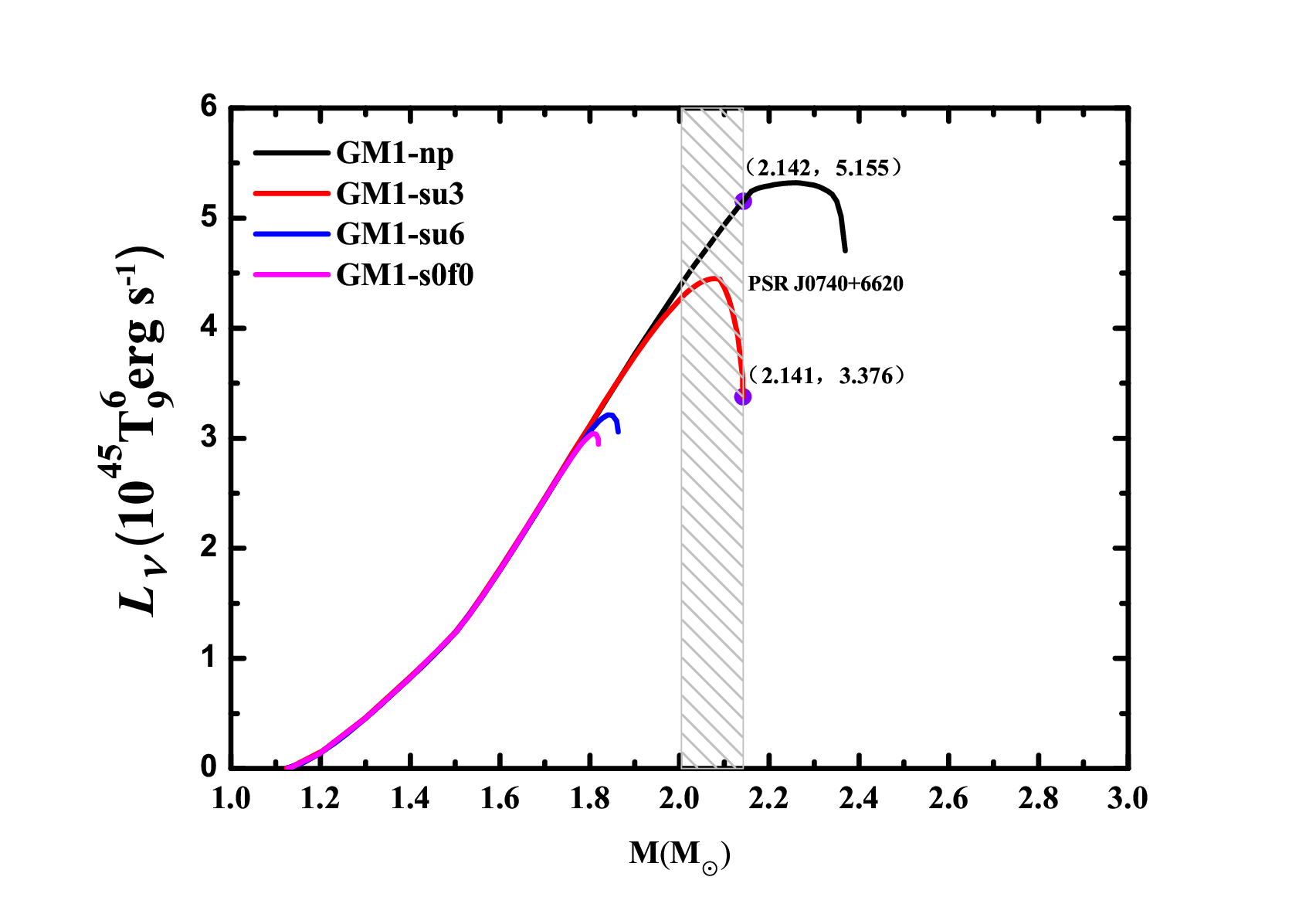}} &
            \subfloat[][\label{f5:b}]{\includegraphics[width=0.45\linewidth]{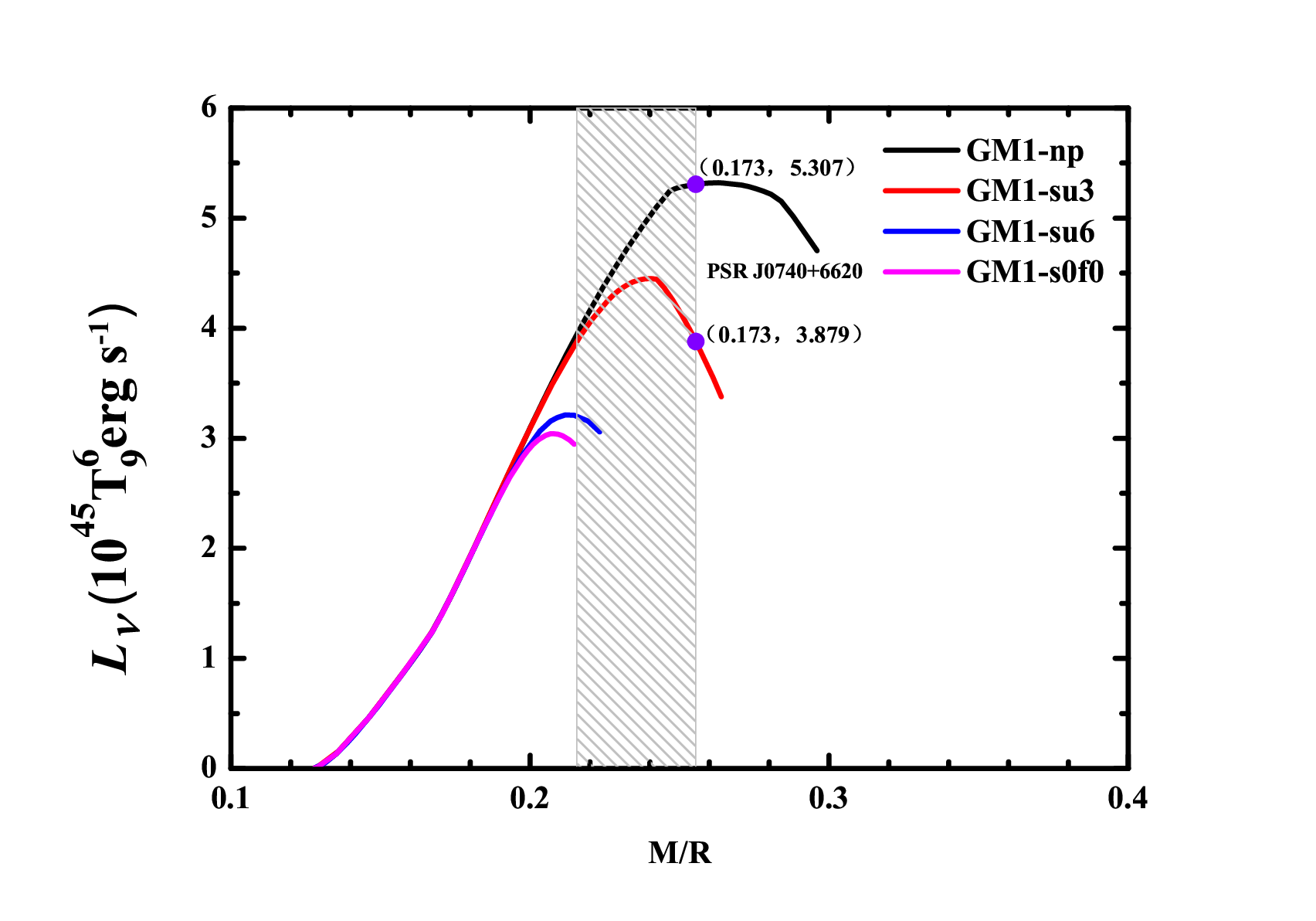}} \\
            \subfloat[][\label{f5:c}]{\includegraphics[width=0.45\linewidth]{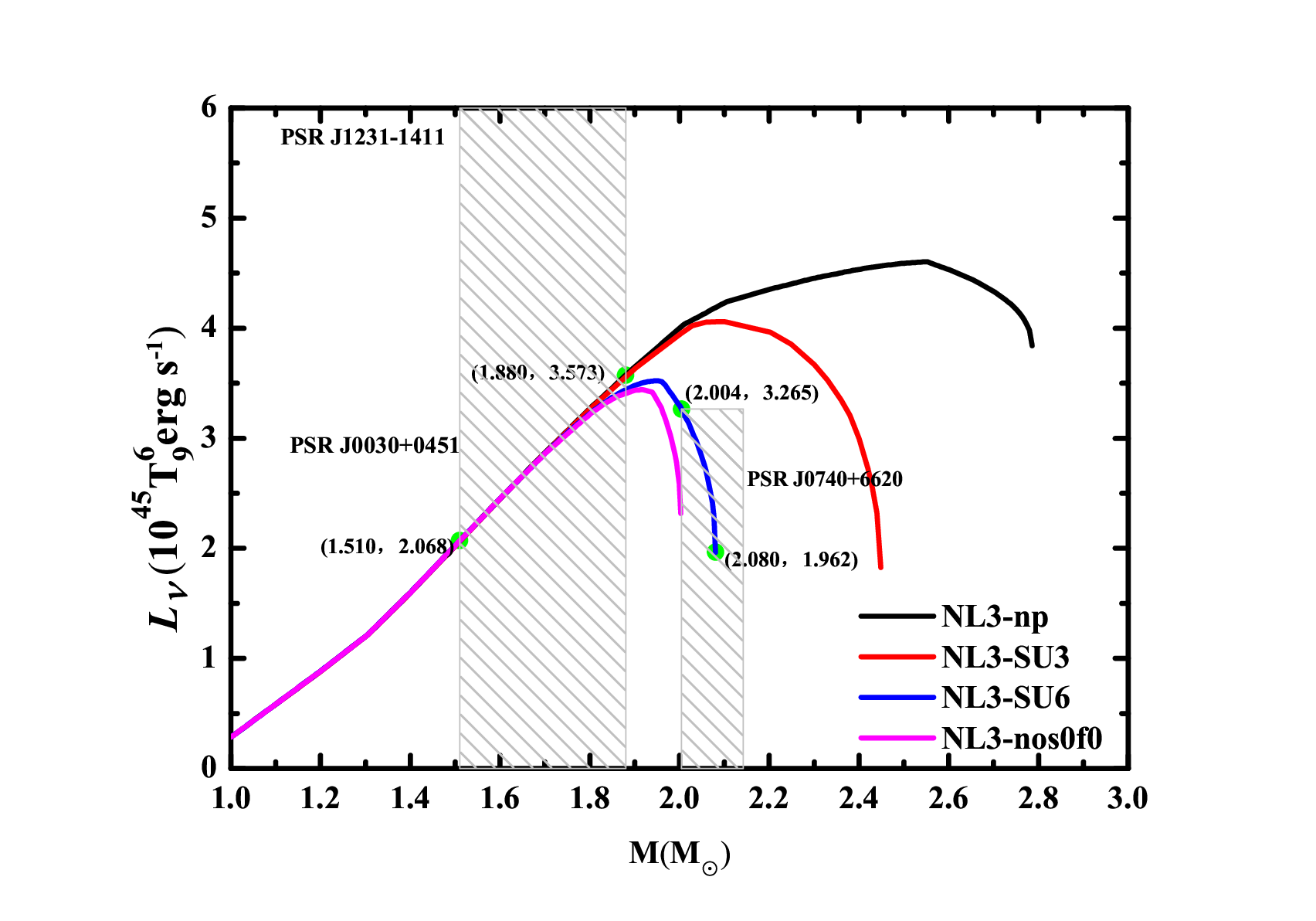}} &
            \subfloat[][\label{f5:d}]{\includegraphics[width=0.45\linewidth]{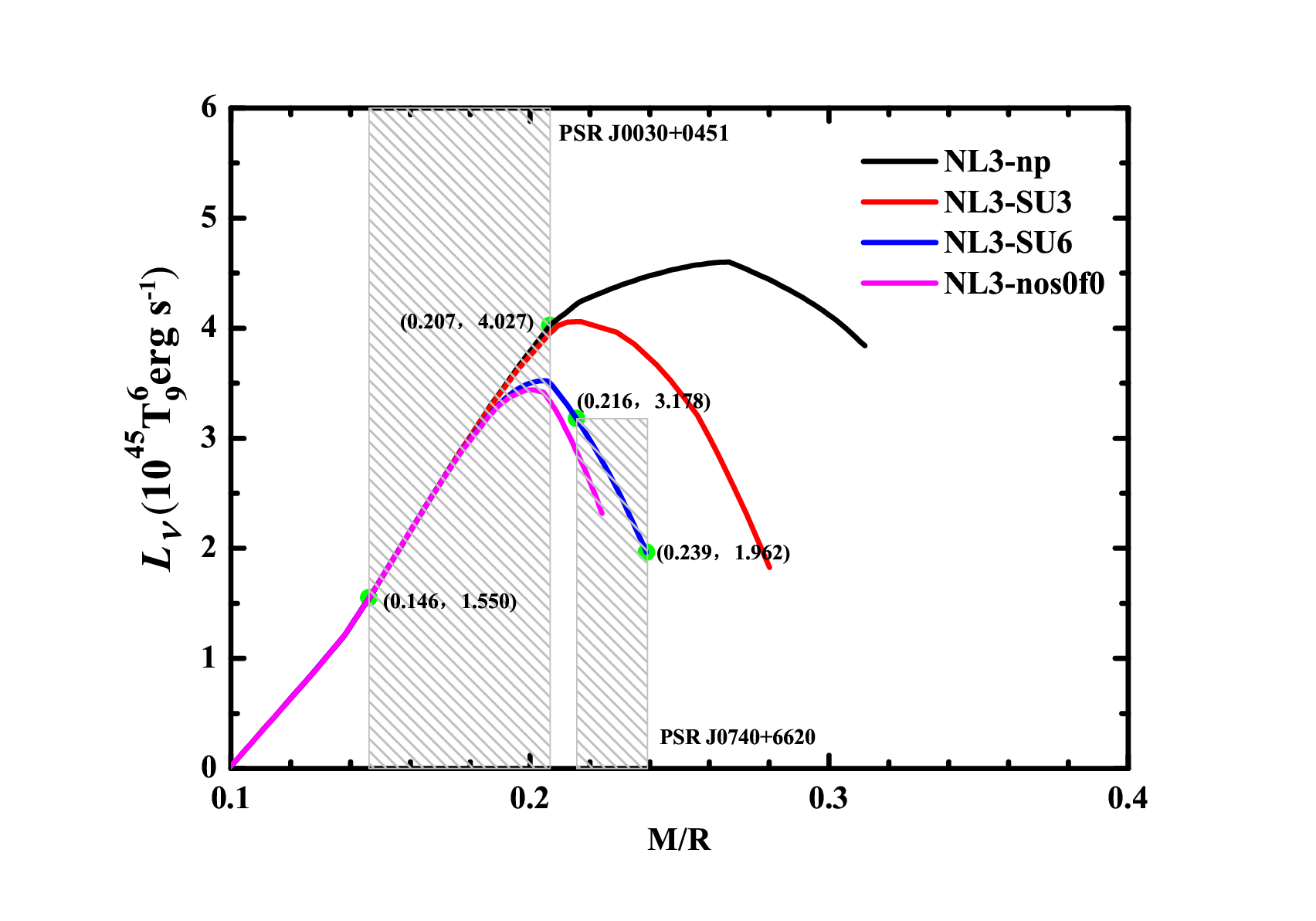}}
        \end{tabular}
    }
    \caption{Neutrino luminosity-mass and neutrino luminosity-compactness relations for four cases of the GM1 and NL3 sets, with purple and green circles denoting the theoretical prediction ranges of the neutrino luminosities for the nucleonic direct Urca processes in PSRs J0030+0451, and J0740+6620.}
    \label{fig:neutrino luminosity}
\end{figure*}

The neutrino luminosities of the nucleonic direct Urca processes, which are functions of the stellar mass and compactness (the mass-to-radius ratio), for the four cases under the GM1 and TM1 parameter sets are plotted in Figure \ref{fig:neutrino luminosity}. As shown in Figure \ref{fig:neutrino luminosity}, regardless of whether hyperons are present in the core of a neutron star, the neutrino luminosities from the nucleonic direct Urca processes in neutron star display a non-monotonic variation trend as the stellar mass or compactness increases. Under the GM1 and NL3 parameter sets, the possible values and ranges of the neutrino luminosity of the nucleonic direct Urca processes in PSRs J0030+0451 and J0740+6620 are shown in Table 2. Under the NL3 parameter set, the $L_{\nu}^{D}$ of PSR J0030+0451 in case (ii) is almost the same as that in case (i), and it is not affected by the hyperonic fraction. However, compared with case (i), the $L_{\nu}^{D}$ of this star in case (iii) or case (iv) is obviously suppressed by the hyperonic fraction when $M_{mv} \geq $ 1.8 $M_{\odot}$ or $(M/R)_{mv} \geq $ 0.18. For the heavier pulsar PSR J0740+6620 in case (ii) under the GM1 parameter set, as the values of $M_{mv}$ or $(M/R)_{mv}$ increase, the hyperonic fraction significantly rises, which greatly suppresses $L_{\nu}^{D}$. This leads to a gradual expansion of the difference in $L_{\nu}^{D}$ between case (i) and case (ii). For this pulsar in case (iii) under the NL3 parameter set, as the values of $M_{mv}$ or $(M/R)_{mv}$ increase, $L_{\nu}^{D}$ shows a rapid decreasing trend. It should be emphasized that all theoretical value ranges presented in this study are based on the selected models and parameters, and must be systematically validated and refined through the accumulation and analysis of future astronomical observational data. If we can obtain relevant observational information about the neutrino emissivity or luminosity of pulsars in the future with the advancement of the astronomical observations, our results offer a possible perspective for judging whether a pulsar contains hyperons and hyperonic species as well as constraining on nuclear physics parameters.

\section{Conclusion}
In this work, we mainly investigate the potential effects of hyperons on the neutrino emissivity and luminosity in neutron stars using the GM1 and NL3 parameter sets under the two ﬂavor symmetries of SU(6) and SU(3) in the framework of the mean ﬁeld theory. We analyzed the feasibility of the nucleonic direct Urca processes and their associated neutrino emission properties focusing on three representative PSRs J1231-1411, J0030+0451 and J0740+6620. Under the GM1 and NL3 parameter sets, the nucleonic direct Urca processes will not occur inside PSR J1231-1411 due to the non-satisfaction of the momentum conservation conditions. Considering the factor of hyperons, the nucleonic direct Urca processes with both $e^{-}$ and $\mu^{-}$ will occur in PSR J0030+0451 under the NL3 parameter set. Due to the large span of its mass measurement values, the hyperon abundance fluctuates significantly, which leads to a significant change in the intensity of the neutrino emissivity of the nucleonic direct Urca process inside it. If the inferred mass of PSR J0030+0451 is greater than approximately 1.8 $M_{\odot}$, the neutrino luminosity of its internal nucleonic direct Urca processes under the SU(3) flavor symmetry is almost unchanged compared with the situation of npe$\mu$ matter. However, it shows an obvious hyperon dependence under the SU(6) spin-flavor symmetry. The neutrino luminosity of the nucleonic direct Urca processes inside PSR J0740+6620 decreases faster with the increase of the hyperonic fraction than that in npe$\mu$ matter under the GM1 parameter set, and its luminosity also shows a monotonically decreasing trend under the NL3 parameter set. The above results and analysis demonstrate that the hyperonic fraction exert differential effects on the properties of the neutrino emissivity and luminosity for the nucleonic direct Urca processes at different mass neutron stars. If the future astronomical observations could provide precise measurements of the neutrino emissivity or surface thermal radiation for pulsars, it will be possible for this work not only to help confirm  the hyperonic species in neutron stars but also offer valuable insights for the nuclear physics parameter determinations.

\end{document}